\documentclass[prb,preprint]{revtex4}

\usepackage{graphicx}
\usepackage{amsmath}

\usepackage{epstopdf}

\newcommand{\cm}{\rho} 
\def\rv{{\bf r}}
\def\Rv{{\bf R}}
\def\pv{{\bf p}}
\def\beq{\begin{equation}}
\def\eeq{\end{equation}}

\begin{document}
\title{An elementary exposition of the Efimov effect}
\author{Rajat K. Bhaduri}
\affiliation{Department of Physics and Astronomy, McMaster University, Hamilton, Canada L9H 6T6}

\author{Arindam Chatterjee}
\affiliation{Physikalisches Institut der Universit\"{a}t Bonn, D-53115 Bonn, Germany}

\author{Brandon P. van Zyl}
\affiliation{Department of Physics, St. Francis Xavier University, Antigonish, NS, Canada B2G 2W5}


\begin{abstract}
Two particles that are just shy of binding may develop an infinite
number of shallow bound states when a third particle is added. This
counterintuitive effect
was first predicted by V.\ Efimov for identical bosons interacting with
a short-range pair-wise potential. The Efimov effect
persists for non-identical particles, if
at least two of the three bonds are almost bound. The Efimov effect has
recently been verified experimentally using ultra-cold
atoms. We explain the origin of this effect
using elementary quantum mechanics, and summarize
the experimental evidence for it.
\end{abstract}

\maketitle
\section{Introduction}
Consider three identical bosons (or fermions with spin and isospin) that occupy a spatially symmetric
$s$-state. Efimov predicted in 1970 that the spectrum obeys a geometrical scaling law,
such that the ratio of the successive energy eigenvalues of the system is
a constant.\cite{efimov} This scaling results in an accumulation
of states near zero energy, with the size of the system growing larger by a factor of about $22.7$ for identical bosons for every state approaching zero energy. The number of these three-body bound states is infinite when the
dimer binding is zero. However, the number of
three-body bound states is actually reduced as the two-body interaction is made more attractive.
More remarkable is the fact that the Efimov effect is independent of
the details of
the interaction, for example, whether it is between atoms with a pair-wise van der Waals
interaction, or between nuclei with a nuclear force. The effect persists even with
unequal masses, as long as two of the three dimer bonds have near-zero binding.
These statements will be made more precise in the following.

A rigorous proof of Efimov's prediction followed in 1972.\cite{amado}
Recently, an analytical solution
for three identical bosons exhibiting the Efimov spectrum has been obtained.\cite{gogolin} Efimov examined the triton and
$^{12}$C (the latter as a bound state of three alpha particles) for manifestations of his prediction. However, for nuclear systems, the interaction between two nucleons cannot be adjusted to yield a two-body zero-energy bound
state. Also, the Coulomb potential between protons often restricts the examples for which two of the three
participants are loosely bound neutrons\cite{mazumdar} in the field
of a nuclear core. However, for ultra-cold neutral atoms with tunable
interaction strengths, these drawbacks are absent. The experimental observation of the Efimov effect was first made by Kraemer {\it et al.}\cite{kraemer} with an optically trapped
dilute gas of $^{133}$Cs atoms at 10\,nK. At such low temperatures, thermal motion does not mask quantum effects. Moreover, the two-body interaction between atoms may
be fine-tuned using ``Feshbach resonance.''\cite{fesh, cohen, chin}
Under such conditions, the recombination losses increase sharply due to
the formation of Efimov trimers, giving the experimental signature of the effect. More recently, Barontini {\it et al.}\cite{bar} have found evidence for two kinds of 
Efimov trimers in a mixture of $^{41}$K and $^{87}$Rb; namely 
KKRb and KRbRb. 

Although there are excellent reviews of Efimov physics (see, for example, Ref.~\onlinecite{braaten}), 
beginning graduate or senior undergraduate students unfamiliar
with the quantum three-body problem might find them inaccessible. 
The Efimov effect arises from the
large-distance (asymptotic) behavior of the 
inverse square interaction. The spectrum resulting from such a 
potential is well-known,\cite{coon, griffiths, rajeev} and its implications
for Efimov physics are discussed in
Ref.~\onlinecite{braaten}.

The main goal of this article is to 
derive, using elementary quantum mechanics, the effective inverse square
interaction in coordinate space of the three-body system. To achieve
this goal, we use the separable form of two-body potential,
for which the algebra is tractable. The 
Efimov effect is universal and does not depend on this 
special choice of a separable potential. A choice of the more familiar 
local form for the potential would entail more advanced
aspects of the three-body problem, while not adding to the essential
physics. Our 
hope is that this article can be used as the 
basis for introducing the fascinating
Efimov effect in an advanced undergraduate, or first year graduate course
in quantum mechanics.

The outline of our paper is as follows. In Sec.~II we briefly
review the low-energy scattering of two particles.\cite{pat} 
The rationale for utilizing a 
nonlocal separable two-body potential near a resonance for the analysis of 
the three-body system\cite{fonseca} is introduced at this
stage. Short discussions of the separable potential, and Feshbach
resonance follow. The relevant properties of the inverse
square interaction for an understanding of the Efimov effect are given in 
Sec.~III.  Our discussion of the properties of the inverse square interaction sets the 
stage for the derivation of the Efimov effect in the three-body problem in Sec.~IV. 
Recent experimental evidence for the Efimov effect, which has 
resulted in a renewed interest in this topic,\cite{phys} is briefly
presented in Sec.~V. Our concluding remarks are in Sec.~VI.

\section{The two-body problem}
\subsection {Low energy parameters}
In the elastic scattering of two particles, the initial and the final
energies are the same, and only momentum transfer may take place. The scattering amplitude is said to be ``on the
energy shell.'' The
scattering cross section and other observables may be described in
terms of a phase shift that occurs in each partial wave of the asymptotic (relative) two-body wave 
function. We assume that the
two-body interaction is central, and falls off faster than $r^{-2}$,
where $r$ is the distance between the two particles. If
three or more particles are involved, the situation is not so simple,
even when the particles interact only pair-wise. If a third particle
is present near the pair, it might take away some energy in the
scattering process. The two-body scattering amplitude is then said to
be ``off the
energy shell," and depends on the short-distance behavior of the
two-body wave function.

The exact 
description of the three-body problem is given by three coupled integral
equations involving the two-body off-the-energy-shell scattering 
amplitudes.\cite{fad} 
The Efimov effect in the three-body problem is remarkable
in the sense that its existence is dictated only by the low-energy two-body 
elastic scattering properties, and not by the off-shell 
behavior. The Efimov effect arises as long as the two-body
interaction between at least two of the pairs is tuned to resonance, regardless of
the interaction potential being, for example, local or nonlocal, of zero range or falling off as $r^{-6}$. At resonance, the two-body wave function is just about bound, and
extends over a large distance.  The tuning of the two-body interaction to resonance is all-important in giving rise to the
Efimov effect.

In this paper we choose a
specific form of a nonlocal potential, called a separable potential, 
that gives an analytical form for the two-body wave function, and 
for which the treatment of the three-body Faddeev equations is
simpler.\cite{amado} In the example of the
three-body problem that we discuss in Sec.~IV, we avoid solving the 
Faddeev equations explicitly, while still obtaining the Efimov spectrum. 

For two-body scattering at very low energies, the interaction is 
effective only in 
the relative $s$-state, because the centrifugal barrier in the higher partial
waves suppresses the particles from coming close together. The $s$-wave
scattering amplitude is well-known to be given by
\beq
f_0(k)
=\frac{1}{k} 
\exp(i\delta_0(k)) \sin \delta_0(k) =(k\cot \delta_0(k)-ik)^{-1},
\label{f0}
\eeq
where $\delta_0(k)$ is the $s$-wave phase shift. The differential
cross-section for $s$-wave scattering is isotropic and is given
by $|f_0(k)|^2$. 
The scattering
between the two particles is well-described\cite{bethe, blatt} at
low energies by two
shape-independent parameters, the scattering length $a$, and the 
effective range $r_0$, where 
\beq
k\cot \delta_0(k)=-\frac{1}{a}+\frac{1}{2} r_0 k^2
+ \ldots
\label{ranger}
\eeq

The effective range $r_0$ is a measure of the range of the
two-body potential, and the scattering length is the intercept of the
asymptotic zero-energy $s$-wave wave function. The reader unfamiliar with this description of the two-body problem 
should study Fig.~1 and read Ref.~\onlinecite{pat}, Sec.~II A.

Consider a situation where the attractive potential is not
strong enough to support a two-body bound state, in which case the sign of the
scattering length is negative.\cite{pat} 
As the potential is made more attractive, the scattering
length becomes more and more negative, and goes to $-\infty$
at resonance (that is, a ``bound state'' at $E=0$), and then flips to 
$+\infty$ as the strength of the attractive potential is increased. A further increase in the attractive strength
drives the scattering length to smaller positive values (but
with $a \gg r_0$), resulting in the formation of a dimer.\cite{pat} 
For a zero-range potential, only the first term on the right-hand side of
Eq.~(\ref{ranger}) survives, and it follows from Eq.~(\ref{f0}) that the 
scattering amplitude is given by $f_0=-(1/a+ik)^{-1}$. 
If we extrapolate the wave number to complex values $k=i\kappa_0$ for a
bound state at $k^2=-\kappa_0^2$, we see that the resulting dimer has 
energy $E_b = -\hbar^2/M a^2$, where $M$ is
the mass of each isolated atom. The corresponding wave function for 
$r>r_0$ is given by 
\beq
\psi_0(r)\sim \frac{\exp (-r/a)}{r}. 
\label{wave}
\eeq
For $r\ll a$, Eq.~(\ref{wave})
reduces to $\psi_0 \sim (1/r-1/a)$, showing the connection of the 
intercept to the scattering length. 

\subsection {Separable two-body potential} 
A separable potential is a special form of nonlocal interaction which 
has been extensively used in the three-body problem in nuclear
physics.\cite{mitra, tabakin} 
It results in
considerable simplification of the Faddeev integral equation, which 
Amado and Noble\cite{amado} exploited to give a rigorous treatment of
the Efimov
effect. The work of Amado and Noble was further developed in Ref.~\onlinecite{adhikari}.
In the following we give the Schr\"{o}dinger equation for two particles
interacting by a separable interaction, and consider the scattering of a heavy
particle of mass $M$ interacting with a light one of mass $m$, with
the mass ratio $\cm=M/m$.  The relative energy is given by 
$E_2 ={k^2}/{\nu'}$, where we have set $\hbar=2m=1$ and
$\nu'=\cm/(\cm+1)$. For simplicity, we assume $\cm \gg 1$, and 
take $\nu'=1$. 
The eigenvalue equation (after the center-of-mass has been removed) is
\beq
\label{this1}
(k^2-H_0)|\psi_k\rangle=V|\psi_k\rangle,
\eeq
where $H_0$ is the kinetic energy operator. In {\bf r}-space Eq.~\eqref{this1}
takes the form (where $\rv$ is the relative coordinate between the two particles)
\beq
(k^2+\nabla^2) \psi_k({\bf r})=\!\int \langle{\bf r}|V|{\bf r'}\rangle \psi_k({\bf r'})\,d^3 r',
\eeq 
and in {\bf p}-space Eq.~\eqref{this1} becomes 
\beq
(k^2-p^2)\psi_k({\bf p})=\!\int \langle{\bf p}|V|{\bf p'}\rangle \psi_k({\bf p'})\,
d^3 p'.
\label{scat}
\eeq

A separable potential in Hilbert space may be written as 
$V=-\lambda |g\rangle\langle g|$, where the negative sign is taken for attraction,
and $\lambda > 0$ determines the strength of the potential. In the coordinate representation, the
attractive separable potential in the $s$-state is given by 
$\langle{\bf r}|V|{\bf r'}\rangle=-\lambda g(r) g(r')$, where $g(r)$ 
is taken to be real. When a bound state is present, the Fourier
transform of $g(r)$ is closely related to the bound state wave
function. 
The Schr\"{o}dinger equation in ${\bf p}$-space for
a bound state, $k^2=-\kappa_0^2$, is given by 
\beq
(\kappa_0^2+p^2)\psi_{\kappa_0}({\bf p})=\lambda g(p)\!\int\!g(p')\psi_{\kappa_0}({\bf p}') d^3 p'.
\label{kappa}
\eeq
In Eq.~\eqref{kappa}, 
\beq
g(p)=\!\int\!\langle{\bf p}|{\bf r}\rangle g(r) d^3 r = \frac{1}{(2\pi)^{3/2}}\!\int 
\exp(- i{\bf p}\cdot{\bf r}) g(r) d^3 r.
\eeq
We may write Eq.~(\ref{kappa}) as 
\beq
\psi_{\kappa_0}(p)=\lambda C_{\kappa_0} \frac{g(p)}{(\kappa_0^2 + p^2)},
\label{yama}
\eeq 
where $C_{\kappa_0}=\!\int g(p') \psi_{\kappa_0} (p') d^3 p'$ is a nonzero
constant. If we multiply
both sides of Eq.~\eqref{yama} by $g(p)$, and integrateover $d^3
p$, we obtain the
equation that determines the binding energy $\kappa_0^2$ for a given
potential 
\beq
\lambda\!\int\! \frac{g^2(p)}{(\kappa_0^2+p^2)} d^3 p =1.
\label{two}
\eeq 
In the next section, a similar equation for the three-body bound state
will be obtained.

For explicit calculations we choose the popular Yamaguchi form\cite{yamaguchi}
$g(p)=(p^2+\beta^2)^{-1}$, giving
\beq
g(r) = \frac{1}{(2\pi)^{3/2}}\!\int\!\exp( i{\bf p}\cdot{\bf r}) g(p) d^3p =
\sqrt{\frac {\pi}{2}}\frac{\exp(-\beta r)}{r}.
\label{short}
\eeq
We substitute this choice of $g(p)$ in Eq.~(\ref{yama}), and take its
Fourier transform. After a little algebra, we find that 
\beq
\psi_{\kappa_0}(r)=\lambda C_{\kappa_0} 
\sqrt{\frac {\pi}{2}}\left(\frac{\exp(-\kappa_0 r)}{r}-
\frac{\exp(-\beta r)}{r}\right)
\label{guchi}
\eeq
For small binding, $\kappa_0$ is only slightly greater than zero. We
choose a short range potential, so that $\beta\gg \kappa_0$, and 
asymptotically Eq.~(\ref{guchi}) becomes the same as Eq.~(\ref{wave})
obtained from the zero-range approximation. 
The universal nature of the Efimov effect is independent of the
specific choice of $g(r)$, as long as $g(r)$ is short range. 

\subsection{Feshbach resonance}
The strength of the interaction of a zero-range potential is
controlled by the scattering length $a$. For cold-atoms the 
scattering length may be varied by 
a magnetic field, making use of the Feshbach resonance.\cite{fesh, cohen, chin}
Near a Feshbach resonance, two unbound 
atoms with relative energy $E$ slightly larger than zero 
may scatter off each other through a potential. This scattering process is said to occur in the 
``open channel.'' 
When the same two atoms come close together in a different total
spin state, they may encounter a different potential, and form a bound or 
quasi-bound state at energy $E_{\rm res}>E$.  The process leading to the formation
of a bound or quasi-bound state is called a ``closed''
channel. 
The scattering state at energy $E$ does not
exist in the closed channel, because the closed channel requires $E_{\rm res}>E$ to 
dissociate the atoms.  Nevertheless, there may be interchannel coupling
through a spin-dependent potential that might cause a virtual transition 
between the two channels. 
The two atoms in the open channel have different magnetic moments
than when they are in the quasi-bound state, and hence the energy gap
between the states may be
controlled by the relative Zeeman shifts. As a result, the mixing between the
two states is more pronounced when the Zeeman energy gap is reduced. 
In a single-channel description,\cite{cohen} the two atoms in the
open channel may make a virtual transition to the closed channel state
and back, giving an effective single-channel scattering length in the open
channel that is 
inversely proportional to the energy gap, $\Delta E$, between the two states, where $\Delta E= \Delta \mu (B-B_0)$. 
Here, $\Delta \mu$ is the difference in magnetic moments of the two-atom system in the two channels, and $B_0$ is the magnetic field at which the scattering length diverges
(at resonance) and changes sign. 
When $B$ is very different from the resonance value $B_0$, the effect of the
coupling is small, and the scattering length reduces to its natural
``background'' value $a_{\rm bg}$. More details in the context of
ultra-cold atoms may be found in Refs.~\onlinecite{moer} and \onlinecite{chin}.
\begin{figure}[hbt!]
   {\includegraphics[angle=0, width=96mm]{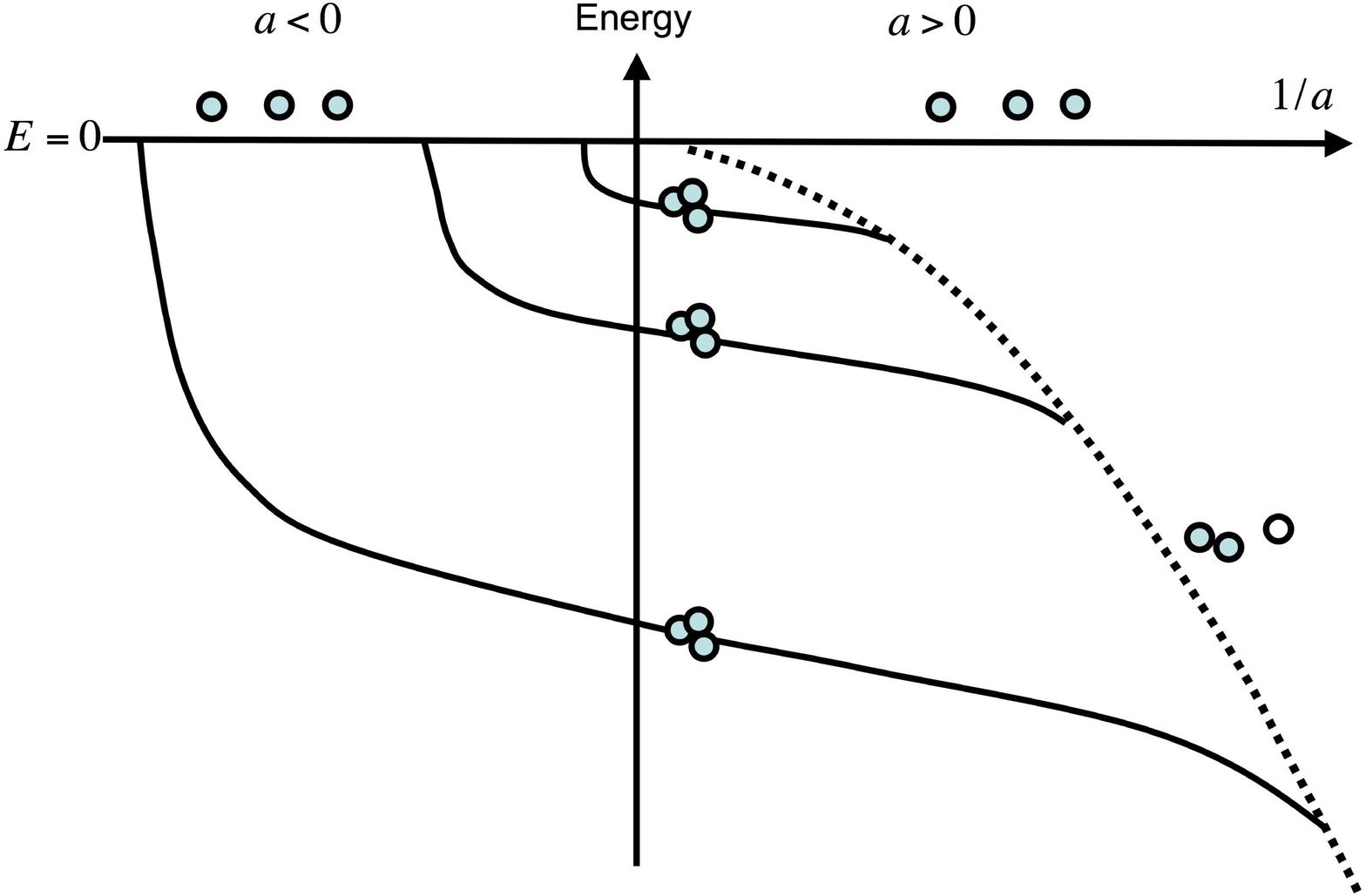}}
   \caption{\label{fig1}%
            Adapted from Fig.~1 in Ref.~\onlinecite{phys}. 
For $a<0$, even though there is no two-body bound state (dimer), more and more
three-body bound states (trimers) are formed as the potential becomes
more attractive. The solid lines represent these Efimov trimer states
which emerge at $E=0$, and $a<0$.  For $a>0$ on the right, the 
trimers break up at the atom-dimer continuum
, indicated by the
  dotted curve}
  \end{figure}
\section{ Efimov spectrum and the Inverse square potential} 
For three identical bosons interacting pairwise via short-range potentials, Efimov predicted
an infinite number of { three-body} bound
states with geometrical scaling, when the two-body interactions are resonant, that is, $|a| \rightarrow \infty$.

Figure 1 (following Ref.~\onlinecite{phys}) depicts the Efimov
scenario where the
energies of some of the three-boson states are plotted as a function of 
$1/a$; for $E>0$ the energies of the three-atoms form a continuum. The bound Efimov states are shown schematically
by solid lines, with a scaling factor (ratio of successive energy eigenvalues) set artificially at 2 rather than $22.7$. The Efimov states break up on the positive side of $a$ in the atom-dimer continuum given by $-\hbar^2/(Ma^2)<E<0$ (see the dotted curve in Fig.~1).

The Efimov spectrum of the three-body problem is the signature of an attractive central potential which falls off asymptotically as the
inverse square power of the distance. A three-body
long-range potential is suggested by noting that for large positive $a$, the size of the
dimer is very large, and the presence of another atom, even if very
distant, may be ``sensed'' by the dimer. For large negative $a$ the two atoms, even
if not bound, are spatially correlated over a distance of order $|a|$ in a quasi-bound state.

To obtain the geometric scaling of the
spectrum, it is sufficient to consider the simpler problem of a single particle of mass $m$ in an inverse square
potential $V(r)=(\hbar^2/2m)\lambda/r^2$, where $\lambda$ is
a dimensionless coupling constant. Classically the equation of motion in
this potential is
scale invariant under the continuous transformations ${\bf r} \rightarrow \alpha {\bf r}$, and $t\rightarrow\alpha^2 t$. Quantum mechanically, for
$\lambda>-1/4$, there is no bound state, and the continuous
scale-invariance is valid. A zero-energy state appears for $\lambda=-1/4$, and the system is anomalous for $\lambda < -1/4$, due to the short-distance
singularity of the potential.\cite{coon,griffiths} A direct
consequence of the anomaly for $\lambda < -1/4$ is that there is no longer a lower limit in
the energy spectrum, and a regularization is required.\cite{rajeev} We are
interested
in the situation where the potential is inverse square only
for $r>r_c$, where $r_c$ is taken as the short-distance
cut-off. We impose the boundary condition that the eigenfunctions vanish at $r=r_c$, which results in a discrete
spectrum. The geometric scaling property, namely that the
ratios of the adjacent energy eigenvalues remain a constant, is
independent of $r_c$.

We write the Schr\"{o}dinger equation in the $s$-state for $r>r_c$ as
\beq
\left[-\frac{d^2}{d r^2}-\frac{(s_0^2+1/4)}{r^2} \right]
u(r)=\frac{2m}{\hbar^2} E u(r),
\label{inverse}
\eeq
where at this stage, $s_0^2\geq 0$ is
just a way of parametrizing the strength of an inverse square potential
that is greater than $1/4$. For bound states, we set $(2m/\hbar^2) E=-\kappa^2$, and require that wave functions vanish
at infinity. We then obtain the solution $u(\kappa r)=\sqrt{\kappa r} K_{is_0}(\kappa r)$, where $K_{is_0}$ is the modified Bessel
function of the third kind of pure imaginary order ${is_0}$.~\cite{braaten}
The boundary condition that $u(\kappa r_c)=0$ makes $\kappa$
discrete, such that $K_{is_o}(\kappa_n r_c)=0$, with $n$ a positive integer. For shallow bound
states such that $(\kappa_n r_c\ll 1)$, the zeros of the Bessel
function $K_{is_0}(\kappa_n r_c)$ are given by
\beq
\kappa_n r_c=\exp\Big(\frac{-n\pi}{s_0}\Big)(2 e^{-\gamma})[1+O(s_0)+\ldots],
\eeq
where $\gamma$ is Euler's constant. 
Equation (14) leads to the desired result
\beq
\frac{E_{n+1}}{E_{n}}= \exp (-2\pi/s_0), \hspace{1cm} n=1,2, \ldots\infty
\label{scaling}
\eeq
which is the geometric scaling mentioned previously. Note that the actual value of $E_n$
scales as $r_c^{-2}$, but geometric scaling holds for the shallow states. Also note that as $n$
becomes larger, the states become shallower, with an infinite number of
states accumulating at zero energy. In the three-body problem in which three particles interact pair-wise,
there are six degrees of freedom after the
center-of-mass motion is eliminated. This problem is commonly treated in
hyperspherical coordinates\cite{nielsen} with a hyperradial variable
$R$, and five angles. For equal mass particles,\cite{note} $R=\sqrt{(r_{12}^2+r_{23}^2+r_{31}^2)/3}$. In the adiabatic
approximation\cite{macek}
for fixed $R$, we solve the Schr\"{o}dinger equation
for the angular variables, thus obtaining a complete set of adiabatic eigenvalues $\varepsilon (R)$ and corresponding eigenfunctions. The solution shows that in the resonant limit $a\rightarrow
\pm\infty$, neglecting channel coupling, the same
inverse square potential as in Eq.~(1) appears in hyperspherical
coordinates, with $r$ replaced by $R$.

For identical bosons, $\exp (\pi/s_0)\simeq 22.694$, (that is, $s_0 \approx 1.00624$) but in general, $s_0$ depends on the mass
ratios. If $a$ is finite and very large, with $|a| \gg r_0$, the inverse square interaction $\varepsilon
(R)$ is cut-off at a short distance of the order of $r_0$, and at a
long distance of the order of $|a|$. The number of shallow bound states is given approximately by\cite{fonseca}
\beq
N \simeq \frac{s_0}{\pi} \ln \frac{|a|}{r_0}.
\label{tom}
\eeq
A simplified derivation of Eq.~(\ref{tom}) is given in the
Appendix.

\section{The three-body model}
Our objective of obtaining an inverse square interaction in the
three-body problem is best served by taking two identical heavy particles
$1$ and $2$, each of mass $M$, and particle $3$ of mass
$m \ll M$, as shown in Fig.~2.
\begin{figure}[hbt!]
   {\includegraphics[angle=0, width=86mm]{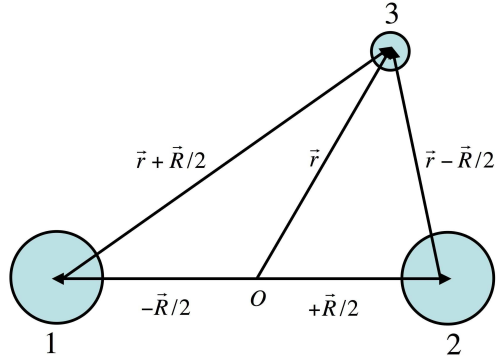}}
   \caption{\label{fig2}%
            A schematic illustration of the relative coordinates used in the three-body problem.  See the text for details.}
\end{figure}

The coordinates of the
particles are labeled $\rv_1$, $\rv_2$, and
$\rv_3$, as measured from an arbitrary origin not shown in Fig.~2. The three-body analysis is simplified if the relative coordinates
$\Rv=(\rv_1-\rv_2)$, and $\rv=\rv_3-(\rv_1+\rv_2)/2$ are introduced. Following the common convention, we denote by $V_1$ the interparticle potential between particles $2$ and
$3$, and likewise for $V_2$ and $V_3$ in cyclic order. The three-body Schr\"odinger equation is given by ($\cm=M/m$, and recall that $\hbar=2m=1$)
\beq
H \Psi (\rv, \Rv) = E \Psi(\rv, \Rv),
\label{muon}
\eeq
with
\beq
H = -\frac{1}{\mu} \nabla^2_R-\frac{1}{\nu}\nabla^2_r +V_1+V_2+V_3, \quad \mu=\cm/2, \quad \nu=2\cm/(2\cm+1),
\eeq
where $E$ is the energy of the three-body system. For $\cm \gg 1$, $\nu \rightarrow 1$, and the motion of the heavy particles is very slow compared to that of the light particle of mass $m$, which is in the spirit of the Born-Oppenheimer approximation. Following Ref.~\onlinecite{fonseca} we may apply the adiabatic Born-Oppenheimer approximation to
solve the three-body Scho\"odinger equation in two stages. First, the wave function is decomposed according to
\beq
\Psi(\rv, \Rv)=\psi(\rv, \Rv) \phi(\Rv),
\eeq
where $\psi(\rv,\Rv)$, with eigenenergy $\varepsilon(R)$, is first solved for the relative motion of the light-heavy system,
keeping $\Rv$ a parameter. For fixed ${\bf R}$, the relative kinetic energy of the heavy particles is zero, and the potential energy $V_3(R)$ is a constant shift in energy. Equation (\ref{muon}), with $\nu=1$, becomes
\beq
\left [-\nabla_r^2 + V_1 (\rv-\Rv/2) +V_2 (\rv +\Rv/2)\right]\psi(\rv,\Rv)=
\varepsilon(R) \psi (\rv, \Rv).
\label{sr}
\eeq
Next, the two-body Schr\"{o}dinger equation with $\varepsilon(R)$ as the
adiabatic potential between the two heavy particles is solved for the ground state energy, $E$, within the adiabatic approximation:
\beq
\left[-\nabla_R^2/\mu +V_3 (R) +\varepsilon(R)\right] \phi(R)= E \phi(R).
\label{heavy}
\eeq
Because the interatomic potential $V_3(R)$ falls off faster than
$1/R^2$, it does not affect the asymptotic behavior of $\varepsilon(R)$,
and we set $V_3(R)=0$. For $V_1$ and $V_2$ in Eq.~(\ref{sr}), we take short-range
separable potentials $-\lambda |g\rangle\langle g|$. Thus, for a fixed
parameter $\Rv$, we have
\beq
\langle \rv'-\Rv/2|V_1|\rv-\Rv/2\rangle=-\lambda \langle \rv'-\Rv/2|g\rangle \langle g|\rv-\Rv/2\rangle,
\eeq
and likewise for $V_2$, with $-\Rv/2$ replaced by
$+\Rv/2$.

If we introduce the displacement operator, $D_{\Rv/2}=\exp\left(i\pv_r\cdot \Rv/2 \right)$, we can write $|\rv+\Rv/2\rangle=D_{\Rv/2}|\rv\rangle$,
where $\pv_r$ is the momentum conjugate to $\rv$. In operator form, Eq.~(\ref{sr}) becomes
\beq
\left[-\nabla_r^2-\lambda \left( D|g\rangle \langle g|D^{-1} + D^{-1}|g\rangle \langle g|D\right)\right]|\psi\rangle =\varepsilon (R)
|\psi\rangle ,
\label{milan}
\eeq
with the subscripts on $D$ suppressed for brevity. Equation~(\ref{milan}) bears a striking similarity to the two-body equation in operator notation given by
$\left[-\nabla_r^2-\lambda |g\rangle \langle g|\right]|\psi\rangle =E_2
|\psi\rangle$. In Eq.~(\ref{milan}), let $\varepsilon(R)=-\kappa^2(R)$, $G=(-\nabla^2_r+\kappa^2)^{-1}$,
and $\langle g|D|\psi\rangle =N(\Rv/2)$. The light-heavy wave function
$\psi(\rv,\Rv)$ has a definite parity under core-exchange, $\Rv\rightarrow -\Rv$, so that for the lowest energy state with energy $E$ [recall that we are solving Eq.~(\ref{heavy})], we take the wave
function to be symmetric. We define
$\langle g|D^{-1}|\psi\rangle\equiv N(-\Rv/2)=N(\Rv/2)$, thereby reducing Eq.~(\ref{milan}) to
\beq
|\psi\rangle =\lambda G \left( D^{-1}|g\rangle + D|g\rangle \right) N(\Rv/2).
\eeq
We multiply both sides by $\langle g|D$, note that $[G,D]=0$ and $N(\Rv/2)\neq 0$, and obtain
\beq
\lambda \left( \frac{\langle g|g\rangle + \langle g|D^2|g\rangle }{(-\nabla_r^2+\kappa^2)} \right)=1.
\eeq
In the momentum representation, Eq.~(20) takes the form
\beq
\label{three}
\lambda\!\int\! \frac{g^2(p_r)}{p^2_r + \kappa^2}d^3p_r +
\lambda\!\int\!\frac{g^2(p_r)\exp(i{\bf p_r}\cdot \Rv)}{p^2_r + \kappa^2}d^3p_r = 1,
\eeq
which is analogous to Eq.~(\ref{two}) for the two-body problem. Note that Eq.~(\ref{three}) assumes $\cm\gg 1$, so that
$\nu'\simeq \nu$.  Recall that $\nu' \equiv \cm/(\cm+1)$ and $\nu=2\cm/(2\cm+1)$.
The coupling constant $\lambda$ may be eliminated by fixing the binding $\kappa_0^2$ of the two-body problem. For the Yamaguchi form $g(p)=(p^2+\beta^2)^{-1}$, the integrals in
Eq.~(\ref{three}) may be performed analytically, giving
\beq
1-\left (\frac{\beta+\kappa_0}{\beta+\kappa}\right )^2= \left (\frac{\beta+\kappa_0}{\beta+\kappa}\right )^2
\left[ \frac{2\beta}{(\beta-\kappa)^2}\frac{e^{-\kappa R}-e^{-\beta R}}{R}
-\frac{\beta+\kappa}{\beta-\kappa} e^{-\beta R} \right ].
\label{four}
\eeq

As the distance $R$ between the two heavy atoms is increased, the
light atom will tend to attach to one of them, and
$\kappa^2(R)\rightarrow \kappa_0^2$. We are particularly interested in
this large $R$ behavior of $\varepsilon (R)$ as the two-body binding $\kappa_0^2$ goes to zero. To deduce the functional form of $\varepsilon
(R)=-\kappa^2(R)$ for large $R$, it is convenient to define $\xi\equiv \kappa -\kappa_0$, and substitute it into Eq.~(\ref{four}). From Eq.~(\ref{short}), we see that $g(r)$ is short-ranged for $\beta \gg 1$. In the resonant limit, consider $a\rightarrow \infty$
from the positive side. Then $\kappa_0=1/a\rightarrow 0$. We consider two possible cases in turn:

(1) $\kappa_0 \ll \beta$, $\beta R \gg 1$, $R/a \rightarrow 0$, $\xi \ll \beta$ for which $\exp (-\beta R) \rightarrow 0$, and $\exp (-\kappa_0 R)\rightarrow 1$. Some algebra yields the
equation
\beq
\frac{e^{-\xi R}}{\xi R}=1,
\eeq
which has the solution $\xi R=A$, where $A=0.5671\ldots$. Hence $\kappa=\kappa_0+\xi= \kappa_0+A/R$. In the limit of
$a\rightarrow\infty$, $\kappa_0\rightarrow 0$, we see that $\varepsilon(R)=-\kappa^2=-A^2/R^2$, giving the desired inverse square
potential. In our analysis, we set $\nu=1$, which only
approximately holds for $\cm \gg 1$. However, setting $\nu=1$ is not necessary, and Fonseca {\it et al.}\cite{fonseca} have shown that $\varepsilon(R)=-A^2/(\nu R^2)$. The effective potential including the kinematic factors that appear in the equation analogous to Eq.~(\ref{inverse}) may be easily deduced
from Eq.~(\ref{heavy}) to be $-\mu A^2/(\nu R^2)=-1/4 A^2(1+2 \cm)/R^2$. Therefore the scaling parameter $s_0$ in the Efimov
spectrum, as defined in Eq.~(\ref{inverse}), depends on the mass ratio $\cm$.

(2) Let the scattering length be large but finite, and consider a distance $R$ that is even larger, so that $R/a \gg 1$. $R/a \gg 1$ implies
that $\kappa_0 R \gg 1$, even though the other conditions are the same as in case (1). Thus, in spite of being able to set $\exp (-\beta R)=0$, we cannot set $\exp(-\kappa_0 R)$ to unity. We may show that
for $R \gg a$, Eq.~(\ref{four}) yields the Yukawa form
\beq
\varepsilon(R) \simeq -\frac{2}{\nu}\frac{e^{-R/a}}{aR}
\label{yu}
\eeq
This form of a static Yukawa potential arises naturally in nuclear
physics due to the exchange of a light mass boson (for example, a pion)
between two heavy mass nucleons. In nuclear physics the range of the potential is
determined by the square root of the mass of the light particle; in
our case, it is the square root of the binding energy of the light
atom that plays the analogous role. This situation is depicted in
Fig.~3. In region I, $R\leq R_0$, where $R_0$ is the range of the interatomic potential\cite{newref} between the two heavy particles. We cutoff the potential for the shallow states at this distance. In region II there is a $1/R^2$ potential, which extends to distances $R \ll a$, hence for all $R$ as $a\rightarrow \infty$, as in case (1). Region III is for $R\sim a$, where the transition to the Yukawa form takes place. Region IV contains the asymptotic behavior of the potential.
\begin{figure}[hbt!]
   {\includegraphics[angle=0, width=86mm]{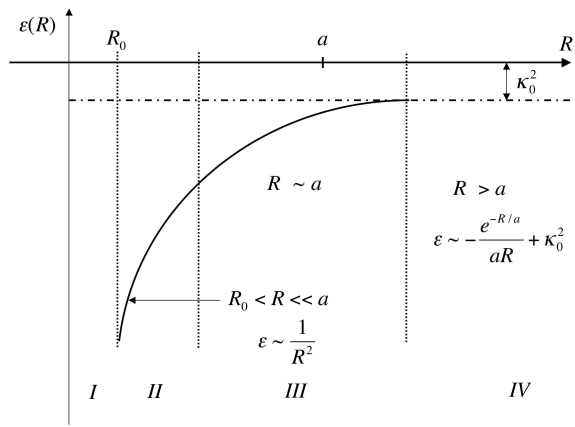}}
   \caption{\label{fig3}%
The effective potential $\varepsilon (R)$ between
     the two heavy-mass particles that arises in the adiabatic 
approximation due to the interaction
     with the light-mass particle. For  
$R<R_0$, the interatomic potential is irrelevant for the Efimov
effect. The scattering length $a$ is marked on the horizontal
axis. As $a\rightarrow \infty$, the two-body binding $\kappa_0^2$ goes
to zero, and $\varepsilon (R)\rightarrow R^{-2}$ for all $R>R_0$.   
The four regions of the potential 
$\varepsilon (R)$, are discussed in the text. }
\end{figure}
As the two-body binding $\kappa_0^2$ goes to zero,
the long-range potential $\varepsilon (R)$ takes the inverse square form, which has no length scale. From
dimensional considerations, because $\hbar^2/M$ has the dimensions of
$E L^2$, an inverse square potential is the only form possible in the
absence of other mass scales or coupling constants.

From Fig.~3, a related prediction associated with the collapse of the
three-body system called the Thomas effect\cite{thomas} may also be
deduced. Consider a two-body system of range $r_0$ with a fixed binding $\kappa_0^2$. Let the range parameter $r_0$ become smaller
and smaller, adjusting the strength of the two-body potential so that $\kappa_0^2$ remains constant. The Thomas effect asserts that the
three-body system will collapse as $r_0\rightarrow 0$, with its
deepest bound state going to $-\infty$. In Fig.~3, the short distance cut-off $R_0$ of the inverse
square potential $\varepsilon(R)$ goes to zero as $r_0$ goes to
zero. Such a behavior near the origin causes a collapse in the
three-body energy. The Thomas effect does not require the asymptotic form of the
potential to be inverse square, but rather is associated with the short
distance singularity of the potential. From Eq.~(\ref{tom}) we note
that the number of three-body bound states diverges when the ratio $a/r_0 \rightarrow \infty$, which may be brought about either by
letting $a\rightarrow \infty$ with $r_0$ finite, as in the Efimov effect,
or letting $r_0 \rightarrow 0$, with $a$ finite, which corresponds to the
Thomas effect.\cite{fred} Unlike the Efimov effect, the Thomas effect is not amenable to experimental verification because
the range of the two-body potential cannot, as of now, be tuned to zero.

\section{Experimental Evidence}
Starting with the pioneering work of Kraemer {\it et al.}\cite{kraemer} with ultra-cold Cs atoms in 2006, several experiments\cite{zacc, gross, pollack} have confirmed Efimov's predictions by measuring the three-body recombination losses
of atoms through the reaction $A+A+A \rightarrow A_2+A$. The experiment with heteronuclear atoms was done by Barontini {\it et al.}\cite{bar}
All these experiments require very low temperature $ k_BT\leq \hbar^2/Ma^2$ for a very large scattering length
$a$ to avoid break up of the dimers. For example, for Cs atoms, $T=10$\,nK was required to largely eliminate thermal effects.

The atoms in a gas have a tendency
to form lower energy dimers directly for scattering length $a>0$. But
two atoms alone cannot form a bound state and preserve momentum and
energy at the same time. Dimer formation is possible if a third atom
is nearby within a distance of the order $a$. This problem was studied
in a gas of identical atoms with number density $n$, while looking at atomic losses in Bose-Einstein condensates.\cite{fed} Let the number of three-body recombinations
per unit volume per unit time be denoted by $\nu_{\rm rec}$, which is proportional to $n^2 (\sigma v) (na^3)$. Here $n^2 (\sigma v)$ is the
probability of two atoms being in the interaction volume $(\sigma v)$, where
$\sigma$ is the elastic scattering cross section and $v$ is the
relative speed between the two atoms. The probability of finding a third atom
within a distance $a$ is $n a^3$. Because $v=\hbar k/m\simeq \hbar/(m a)$, and $\sigma \propto a^2$, we obtain
\beq
\nu_{\rm rec}= C(a) n^3 \left(\frac{\hbar}{m} a^4\right),
\eeq
where $m$ is the mass of each atom, and $C(a)$ is a dimensionless
constant. The variation of $C(a)$ with the scattering length $a$ exhibits the
emergence of an Efimov trimer (negative $a$), and its subsequent
dissolution to dimer plus atom (positive $a$), see Fig. 1. It has been 
calculated from theory.\cite{fed, nielsen2,esry}
The recombination length (the ordinate of Fig.4) is defined as 
$\rho_3=(2\sqrt{3} C(a))^{1/4}a$, where the numerical factor is
included to take into account the reduced mass of the trimer, and the fact that
three atoms are lost for each trimer.
\begin{figure}[hbt!]
   {\includegraphics[angle=0, width=136mm]{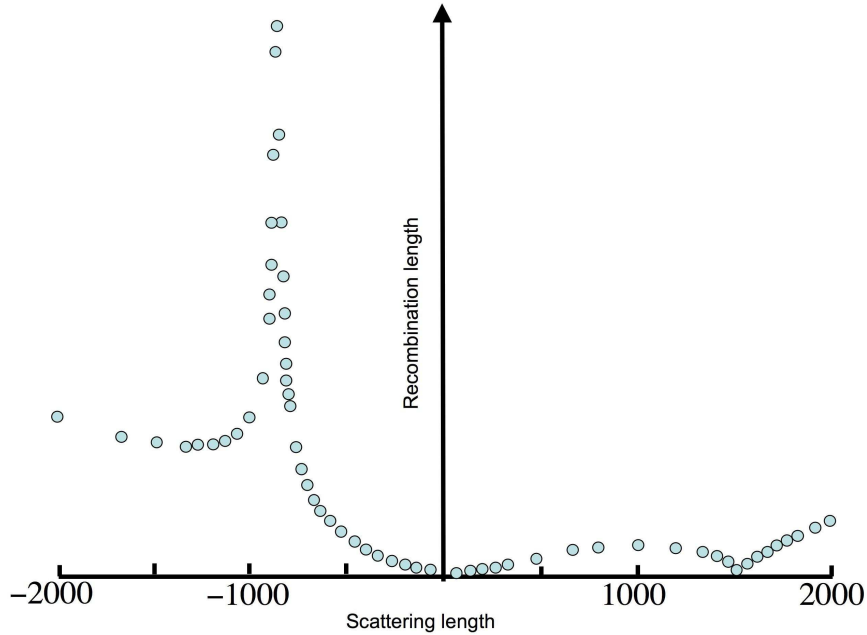}}
   \caption{\label{fig4}%
            Adapted from Fig.~2 of Ref.~\onlinecite{esry2}.  The solid circles
            are experimental data points~\cite{kraemer} at $10$ nK for the
            observed three-body 
recombination: $Cs+Cs+Cs \rightarrow Cs_2 + Cs$.  The scattering
length is measured in units of the Bohr radius $a_0$.}
\end{figure}
To understand the experimental peak in the measured recombination
length $\rho_3$ (see Fig.~4), note that shallow
dimers are not formed directly for $a <0$. The potential
$\varepsilon (R)$ of Fig.~3 develops a potential barrier for
large $R$, followed by an attractive potential at shorter distances. 
Such a potential emerges from a hyperspherical calculation for equal
mass bosons.\cite{macek}
There may result, for certain negative values of $a$, shape resonances 
of the trimer at $E>0$. When the energy of the three atoms matches a
resonance, there is enhanced barrier penetrability to form an Efimov
trimer. The enhanced barrier penetrability allows the system to relax 
to a deeply bound dimer state (from the interatomic potential), with the excess
energy being carried away by the third atom. The peak $\rho_3$ on the negative
side is at $a=-850 a_0$, where $a_0$ is the Bohr radius. The next
peak should be at a value of $a$ that is $22.7$ times larger, that is,
$a=-19,295$, which is outside the range of experimental observation. 
Nevertheless, these multiple peaks were observed in later
experiments.\cite{zacc, gross, pollack}
Even though multiple resonance peaks were not seen in the Kraemer {\it et al.}\cite{kraemer} experiment, the interference minimum 
observed on the positive
$a$ side is convincing evidence for the Efimov effect.\cite{esry2}
Hyperspherical calculations for $a>0$ not only give an attractive
potential as in Fig.~3 for $\varepsilon (R)$, but also a barrier for
positive energies that falls off for large $R$. The three incoming
atoms may be reflected by the barrier, and form $A_2+A$ on the way
out, interfering with the incoming dimer plus atom system following
the large-$R$ attractive channel of Fig.~3. 
The out-of-phase interference between the two paths leads to
the curve that fits the experiment for positive $a$ (see Fig.~2 of
Ref.~\onlinecite{esry2}).

\section{Closing remarks}
It is remarkable and not foreseeable in 1970 that
the Efimov effect has been experimentally verified
using ultra-cold atoms. We have considered a
simple case for which two of the atoms are very heavy compared
to the third. The light heavy pairs are treated very differently, as 
evidenced by the adiabatic approximation. The light particle is almost
unbound, and can be very distant from the heavy partners. 
The overall size of the shallow Efimov trimer is large in
comparison to the short-range of the two-body interaction.

As is clear from Sec.~IV, it is the dynamics of the light particle that 
generates the attractive long-range interaction between the heavy
ones. We may regard the
light particle as being exchanged back and forth to generate this
potential between the two heavy particles. 
Even for equal mass particles, 
Efimov\cite{comments}
has stressed this intuitive picture for the inverse square potential.
A different approach must be 
taken for three identical particles, as indicated briefly in
Sec.~III for three identical bosons. Each particle has to be treated
the same way, and hyperspherical coordinates are well-suited to 
describe the shallow Efimov states. 
The size of these states is large, and
the small binding results in the formation of floppy triangles. For
these shallow states, the 
coupling between the hyperradial variable $R$ and the hyperangles is
small. The coupling vanishes for $|a|\rightarrow \infty$, and the
hyperangular contribution just adds a centrifugal term for angular
motion to the adiabatic potential $V(R)$. The deepest potential is
attractive enough (for $L=0$) to yield the Efimov effect. 

For three nucleons the overall wave function under
the exchange of any two has to be antisymmetric. As was originally found by Efimov,~\cite{efimov}
imposing these symmetries results in only the spatially symmetric angular
momentum state $L=0$ (with the spin-isospin combination antisymmetric) leading
to an attractive interaction. Unlike charge neutral ultra-cold atoms,
it is not possible to manipulate the strength of the nucleon-nucleon interaction.

Given that there is suggestive experimental evidence for the 
Efimov effect in the
four-body sector (see for example, Ref.~\onlinecite{pollack}), it is clear that
Efimov physics in ultra-cold atoms will continue to be an active 
area of research. 

\appendix*
\section{Derivation of bound states in the inverse square potential}
One way to obtain the number of bound states of a potential is to
calculate the canonical partition function $Z(\beta)=\sum_i
\exp(-\beta E_i)$, where $\beta$ for our purpose may be taken to
be a positive parameter. The expression for $Z(\beta)$ may be written as
$Z(\beta)=\!\int_0^{\infty} g(E) \exp(-\beta E) dE$, where $g(E)=
\sum_i \delta(E-E_i)$ is the density of states. The latter
equation shows
that $Z(\beta)$ is the Laplace transform of $g(E)$. The inverse
Laplace transform of $Z(\beta)$ with respect to $E$ yields the
density of states. The integration of the density of states in an energy
interval gives the number of states.
In the following, we present a simple semiclassical derivation of the number of
shallow Efimov states in the energy interval between $0$ and $E$.

Consider the one-body inverse square potential given in Eq.~(1)
\beq
\label{invr2}
V(r)=-\frac{\hbar^2}{2m} \frac{(s_0^2+1/4)}{r^2}.
\eeq
This form of the potential is taken to be valid for $r_0\leq r \leq a$ and for
large $a$ (it is assumed that $r_0\simeq \tilde{r}$). We shall now derive Eq.~(3).
The semiclassical partition function for a given partial wave is
\beq
\label{zl}
Z_\ell(\beta)= \frac{1}{h}\!\int_{-\infty}^{\infty} dp_r \exp(-\beta
p_r^2/2m)\!\int_0^{\infty} dr\exp(-\beta V_\ell(r)).
\eeq
After performing the momentum integral in Eq.~(\ref{zl}), we obtain
\beq
Z_\ell(\beta)=\left(\frac{m}{2\pi\hbar^2}\right)^{1/2}\!\beta^{-1/2} \!\int_0^{\infty} e^{-\beta V_\ell(r)} dr.
\eeq
The Laplace inversion of $Z_\ell(\beta)$ with respect to $\beta$ gives the density of states
\beq
\label{gl}
g_\ell(E)=\left(\frac{m}{2\hbar^2}\right) \frac{1}{\pi}\!\int \frac{dr}{\sqrt{E-V_l(r)}} \Theta(E-V_\ell(r)).
\eeq
For the inverse square interaction a WKB-type approximation yields
exact results when the Langer correction is implemented,\cite{guerin}
that is, $\ell(\ell+1)$ is replaced by $(\ell+1/2)^2$. Hence,
\beq
V_\ell(r)=-\frac{\hbar^2}{2m} \frac{(s_0^2+1/4)}{r^2}+\frac{\hbar^2}{2m r^2}
(\ell+1/2)^2.
\eeq
We are interested in the $\ell=0$ partial wave, for which
\beq
V_0(r)=-\frac{\hbar^2}{2m} \frac{s_0^2}{r^2}.
\label{v0}
\eeq
The number of states is obtained by integrating Eq.~(\ref{gl}) with respect
to $E$, followed by the integration with respect to $r$:
\beq
N(E)= \left(\frac{m}{2 \hbar^2}\right)\frac{2}{\pi} \int_{r_0}^a \sqrt{E-V_0(r)} dr.
\eeq
If we take $E$ close to zero and $V_0(r)$ from Eq.~(\ref{v0}, we obtain
\beq
\label{neffimov}
N \simeq \frac{s_0}{\pi} \ln \Big(\frac{a}{r_0}\Big),
\eeq
which is the well known result for the number of shallow Efimov states.\cite{efimov}
The same derivation applies to the
three-body problem with potential $\varepsilon (R)$ (see Fig.~3), when, for
very large $a$, we take the form Eq.~(\ref{invr2}) with $R_0\leq R\leq a$.

\begin{acknowledgments}
RKB and BPvZ would like to thank the Natural Sciences and Engineering Research Council
of Canada (NSERC) for financial support under the Discovery Grants
Program. We would like
to thank D.\ W.\ L.\ Sprung and Akira Suzuki for carefully reading through the manuscript, along with the anonymous referees for their helpful suggestions.
\end{acknowledgments}


\begin{thebibliography}{99}
\bibitem{note} This definition is taken from Ref.~\onlinecite{braaten}. Many authors, including Efimov\cite{efimov} define $R^2=\frac{2}{3} (r_1^2+r_2^2+r_3^2)$.
\end{thebibliography}
\end{document}